\documentclass[11pt]{article}
\pdfoutput=1
\usepackage{jcappub,natbib,float,verbatim}
\usepackage{color}
\usepackage{comment,subcaption,multirow}
\usepackage{tikz,booktabs,makecell}
\newcommand{\head}[1]{\textnormal{\textbf{#1}}}
\title{Time-delay Cosmography: Increased Leverage with Angular Diameter Distances}
\author{I. Jee$^{a}$, E. Komatsu$^{a,b}$, S. H. Suyu$^{c,a}$, D. Huterer$^{d,a,e}$}
\affiliation[a]{Max-Planck-Institut f\"ur Astrophysik, Karl-Schwarzschild Str. 1, 85741 Garching, Germany}
\affiliation[b]{Kavli Institute for the Physics and Mathematics of the
Universe, Todai Institutes for Advanced Study, the University of Tokyo,
Kashiwa, Japan 277-8583 (Kavli IPMU, WPI)}
\affiliation[c]{Institute of Astronomy and Astrophysics, Academia Sinica,
P.O. Box 23-141, Taipei 10617, Taiwan} 
\affiliation[d]{Department of Physics, University of Michigan, 450 Church St, Ann Arbor, MI 48109-1040} 
\affiliation[e]{Excellence Cluster Universe, Boltzmannstrasse 2, D-85748 Garching, Germany}
\emailAdd{ijee@mpa-garching.mpg.de}
\abstract{%
Strong lensing time-delay systems constrain cosmological parameters via the so-called time-delay distance and the angular diameter distance to the lens. In previous studies, only the former information was used in forecasting cosmographic constraints. In this paper, we show that 
the cosmological constraints improve significantly when the latter information is also included. Specifically, the angular diameter distance plays a crucial role in breaking the degeneracy between
the curvature of the Universe and the time-varying equation of state of dark energy. Using a mock sample of 55 bright quadruple lens systems based on expectations for ongoing/future imaging surveys, we find that adding the angular diameter distance information to the time-delay distance information and the cosmic microwave background data of Planck improves the constraint on the constant equation of state by 30\%, on the time variation in the equation of state by a factor of two, and on the Hubble constant in the flat $\Lambda$CDM model  by a factor of two. Therefore, previous forecasts for the statistical power of time-delay systems were significantly underestimated, i.e., time-delay systems are more powerful than previously appreciated.
}
\begin{document}
\maketitle
\flushbottom
\section{Introduction}
%
The redshift-distance relation constrains cosmological parameters.
The use of strong gravitational lens systems with time-delay measurements 
as a cosmological distance probe was suggested by Refsdal in 1964 \citep{refsdal:1964}, 
and has been extensively studied to measure the present value of the Hubble parameter, $H_0$ (see, e.g. \citep{suyu/others:2010}).
The physical quantity of interest here is the so-called \textit{time-delay distance} $D_{\Delta t}$.
This is a distance-like quantity, given by a combination of three angular diameter distances in a strong lens system:
\begin{equation}
D_{\Delta t} \equiv (1+z_L)\frac{D_\mathrm{A}(EL) D_\mathrm{A}(ES)}{D_\mathrm{A}(LS)},
\end{equation}
where $z_L$ is the redshift of the lens, while $D_\mathrm{A}(EL)$, $D_\mathrm{A}(ES)$ and $D_\mathrm{A}(LS)$ are the angular diameter distances 
from the Earth to the lens, from the Earth to the source, and from the lens to the source, respectively. 
As each of the distance is proportional to the inverse of $H_0$, $D_{\Delta t}$ is 
proportional to $1/H_0$. Ref. \citep{hu:2005} has shown that a precise measurement of $H_0$ 
is important for constraining the equation-of-state parameter of dark energy, $w$. This motivated the measurement of time-delay distances in the recent years. 
However, the constraints weaken significantly when $w$ is allowed to vary with time. 

It is, in fact, possible to extract the angular diameter distance to the lens, $D_\mathrm{A}(EL)$, 
by combining time-delay measurements with the lens stellar velocity dispersion measurements \citep{paraficz/hjorth:2009, jee/komatsu/suyu:2014}. The physics is simple:
The time-delay measurement is the mass estimate of the lens galaxy, while the velocity dispersion measurement
is the potential estimate. By knowing the mass and the potential, we can calculate the physical size of the
system, thus the system can be used as the ``ruler''. Comparing the observed separation of lensed images with the length of the ruler, we can estimate $D_A(EL)$.
In ref. \citep{paraficz/hjorth:2009}, Paraficz and Hjorth demonstrated this for a 
singular isothermal sphere (SIS) lens without mass external to the lens along the line of sight. 
In ref. \citep{jee/komatsu/suyu:2014}, we have extended the analysis by including the external convergence, 
allowing the mass profile of the lens to follow an arbitrary power-law, and allowing the velocity dispersion to be anisotropic.
We have found that the main source of uncertainty for $D_A$ is the unknown (anisotropic) velocity structure, 
which affects the normalization of the potential of the lens galaxy. 
We have also found that the mass external to the lens along the line of sight (external convergence) 
has no effect on the inferred $D_A(EL)$.
In this paper, we show that the constraints on the cosmological parameters, especially on a time-varying $w$,
improve significantly by including $D_A(EL)$.

The structure of the paper is as follows. In section \ref{sec:method}, we introduce the mock catalog of the
strong lens systems we use, and describe the cosmological model along with the fiducial cosmological parameters we use. 
In section \ref{sec:single_probe}, we compare the cosmological constraints we expect from lenses to 
those from the other cosmological distance probes, specifically, Type Ia Supernovae (SNe) and Baryon Acoustic Oscillation (BAO). In section \ref{sec:pivot} we introduce the pivot redshift and $w_p$-$w_a$ parametrization, and show how lensing distances improve the figure of merit for the dark energy equation of state. We combine cosmological 
information from different probes to show the constraining 
power of strong lenses in practice, and conclude in section \ref{sec:conclusion}. In Appendix \ref{sec:lens_on_h}, we show the constraint on the Hubble constant $H_0$ using the two lensing distances. In Appendix \ref{sec:flat_univ}, we show the dark energy constraints assuming that the universe is spatially flat.
\section{Method}
\label{sec:method}
Each well-modeled time-delay lens system yields two distance(-like) quantities, $D_A(EL)$ and $D_{\Delta t}$. 
The uncertainties of $D_A(EL)$ and $D_{\Delta t}$ are dominated by the velocity dispersion 
and the external convergence, respectively. In this work we make an assumption that 
we can measure both distances
with $5\%$ uncertainty, which requires a few per cent measurement of the spatially resolved velocity dispersion of the lens galaxy, as well as a good understanding of the mass distribution along the line-of-sight, that is obtainable by simulations and observations of the lens environment \citep{suyu/others:2013,jee/komatsu/suyu:2014}. Regarding the lens mass model, the power-law density profile in ref. \citep{jee/komatsu/suyu:2014} is widely used due to its ability to fit the imaging data near the image positions. The local density profile is well reconstructed with the model if the images are spatially extended such that information from thousands of intensity pixels can be used. However, ref. \citep{schneider/sluse:2013a,schneider/sluse:2013b} have pointed out that the information obtained by the lensed images cannot uniquely determine the shape of lens mass profile due to the so-called \textit{Source-Position Transformation} (SPT). Specifically, they focused on the degeneracy between composite density profiles and a power-law mass profile, and have shown that fixing the shape of the lens mass profile as a power law can break the SPT. However, they have also mentioned that these models can be distinguished if more information is available: for example, if more than three images with time delays are observed, the degeneracy can be broken as the general SPT does not conserve the time delay ratios. In ref. \citep{suyu/others:2013}, the robustness of the measured time-delay distance is tested with power law and composite model under the presence of lens kinematics data and shown to be nearly independent of the choice of the model. Ref. \citep{falco/others:1985} has shown that the so-called \textit{Mass-Sheet Transformation} (MST), which is a special case of the SPT, scales the time delays by the same factor, and thus conserves the time-delay ratio, can bias the mass modeling. However, ref. \citep{xu/etal:2015} has empirically shown using the Illustris simulation that most of the early-type galaxies with high ($\sigma>200  $kms$^{-1}$) velocity dispersion, which most of the lens galaxies are, show nearly power-law behavior. The MST degeneracy can also be broken if additional information on the lens galaxy (e.g. velocity dispersion) is obtained. Ref. \citep{keeton/moustakas:2008} has shown that the existence of substructures in the lens galaxy perturbs the time delay: However, the effect of perturbation on the time delays ($<$1 day) is typically smaller than the currently available time-delay  measurement uncertainties, and thus both distances are mainly determined by the global mass distribution rather than the substructures. Thus we claim that this precision measurements on both distances is possible, but only when good quality imaging / kinematics data as well as time-delay measurements are available. The correlation between the two distances is negligible, because the uncertainties in the velocity dispersion and the external convergence
are uncorrelated.

To study the expected cosmological constraints from lenses, we need to specify the distribution
of lens and source redshifts. We use the catalog of mock lenses in ref. \citep{oguri/marshall:2010} to obtain 
the redshift distribution of time-delay lenses with double and quadruple images expected for the 
Large Synoptic Survey Telescope (LSST) \citep{abell/etal:2009, ivezic/etal:2008}. 
Although LSST itself is expected to find around ten thousand lensed quasars, 
there are only about 400 systems that would have good time delay measurements \citep{liao/etal:2014}.
To obtain distances from a lens system with a reasonable accuracy, a good mass model of the lens galaxy 
is also required, as both the time-delay distance and the angular diameter distance are sensitive to 
the mass distribution of the lens. 
Ancillary data both in terms of high-resolution imaging and spectroscopy of the lens systems are needed for accurate lens mass modeling. Therefore, we select lens systems from the mock catalog with the following criteria for acquiring ancillary data with relative ease: (1) the quasar image separation is $>1''$, (2) the third brightest quasar image has an $i$-band magnitude $m_i<21$, and 
(3) the lens galaxy has $m_i<22$.  The criteria on the quasar image separation and brightness make it easier to measure the time delays with high precision (uncertainty of a few percent).  Furthermore, a sufficiently wide quasar image separation is required for extracting the Einstein ring of the quasar host galaxy and measuring the lens velocity dispersion for mass modeling. The lens galaxy also needs to be of sufficient brightness for measuring the lens velocity dispersion.  In this work, we focus on quadruply imaged lens systems as they provide more information than doubly imaged systems.  After applying these criteria to the LSST mock lens sample, we obtain 55 quadruple lens systems as the best cases of obtaining ancillary data. 

\begin{figure}[H]
\begin{center}
\includegraphics[height=9cm]{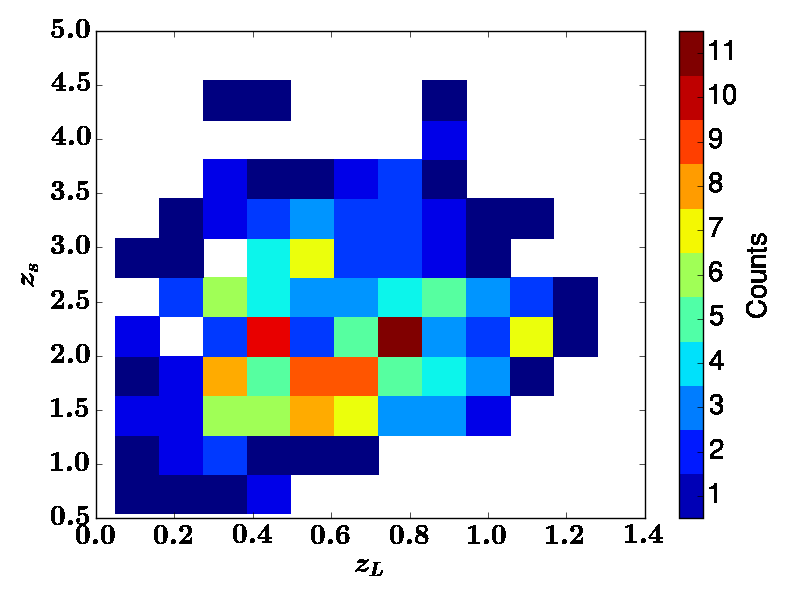}
\caption{The distribution of source ($z_S$) and lens ($z_L$) redshifts of quadruply imaged time-delay lenses expected for LSST \citep{oguri/marshall:2010}. These lens systems are the best ones to obtain ancillary data and measure  distances with 5\% precision. For visualization purposes, the number of the lenses is oversampled by a factor of five only in this figure (thus each pixel does not necessarily have a multiple-of-5 value).}
\label{fig:quads}
\end{center}
\end{figure}

Figure \ref{fig:quads} shows the source and the lens redshift distribution of quadruple lenses 
in our mock catalog. For the visualization purpose, only in this figure, the total number of detectable lenses is 
oversampled by a factor of 5 (based on the catalog from ref. \citep{oguri/marshall:2010}) to populate the histogram. The expected constraints reported in this paper are derived from
the actual distribution of the 55 lenses.

Since these are the bright lens systems, a fraction of these systems will already be discovered in the current imaging surveys.  In particular, we expect that $\sim$25\% of these systems will be discovered in the Dark Energy Survey (DES)\footnote{http://www.darkenergysurvey.org/index.shtml} and the Hyper Suprime-Cam (HSC; \citep{miyazaki/etal:2012})\footnote{http://www.naoj.org/Projects/HSC/surveyplan.html} Survey.  Furthermore, we expect a few more quadruple lens systems from the northern areas of the HSC Survey that are not covered in DES and LSST.  Therefore, even though we focus here on the LSST sample, our cosmographic predictions are also relevant for the upcoming years before the LSST era as new lens systems in the current imaging surveys are discovered and monitored.

We explore constraints on two variations of $\Lambda$CDM model. 
Both assume a curved universe ($\Omega_k \neq 0$) and 
an unknown equation of state of dark energy ($w \neq -1$). 
The first model assumes that $w$ is a constant (o$w$CDM) with the following cosmological parameters:
\begin{equation}
\vec{\theta} \in \{\Omega_m, \Omega_{k}, w, h\} \quad \mbox{(o$w$CDM model).}
\end{equation}
The second model further assumes that $w$ varies in time with
$w = w_0 + (1-a)w_a$ \citep{linder:2003} (o$w_\mathrm{z}$CDM):
\begin{equation}
\vec{\theta} \in \{\Omega_m, \Omega_{k}, w_0, w_a, h\} \quad \mbox{(o$w_z$CDM model).}
\end{equation}
We choose the fiducial cosmology following Planck 2015
($\Omega_{m} =0.308$, $\Omega_{k} = 0$, $w_{0} = -1$, $w_{a} = 0$, and $h = 0.678$) \citep{ade/etal:2015b}. 

We use the Fisher information matrix (hereafter the Fisher matrix) to calculate the constraining power of the cosmological probes. 
For a data vector $\vec{d}(\vec{\theta})$ with a set of parameters $\vec{\theta}$, the 
Fisher matrix {\bf F} is given by
\begin{equation}
F_{ij} = \sum_{\alpha \beta} \frac{\partial d_{\alpha}}{\partial \theta_i}~{\bf Cov}^{-1}_{\alpha \beta}~\frac{\partial d_{\beta}}{\partial \theta_j}~,
\label{eq:fisher}
\end{equation}
where indices $\alpha$ and $\beta$ run over the observables, and ${\bf Cov}(\vec{d}$) is the data covariance matrix. 
For lenses, ${\bf Cov}_{\alpha \beta} = \delta_{\alpha \beta} \sigma_{\alpha}^{-2}$,
where $\delta_{\alpha \beta}$ is the Kronecker delta as we assume no correlation
between the different lens systems and between the two measured lensing distances $D_A$ and $D_{\Delta t}$ of each lens system. The uncertainty in each distance $\sigma_{\alpha}$ is $\sigma_{\alpha} = 0.05 d_{\alpha}$ as we assume 5\% precision measurements of both distances.
The inverse of the Fisher matrix, ${\bf F}^{-1}$,
gives the covariance matrix of the parameters, and the marginalized uncertainty on the \textit{i}-th parameter is calculated as $({\bf F}^{-1})_{ii}^{1/2}$. If the normalization of {\bf F} increases by a factor of $n$, then
the normalization of the parameter covariance matrix decreases by a factor of $n$, thus the error bar on each parameter
tightens by a factor of $\sqrt{n}$. 
%

\section{Single-probe constraints combined with the Planck distance prior}
\label{sec:single_probe}
In this section we investigate the constraining power of the time-delay lenses expected from 
LSST in section \ref{sec:lens}, and compare the constraints to that of two other cosmological distance probes, 
the BAO data from Baryon Oscillation Spectroscopic Survey (BOSS) data release (DR) 11 in section \ref{sec:bao},
and the SNe data from Joint Light-curve Analysis (JLA) in section \ref{sec:sne}. We examine the future prospects of constraints from BAO and SNe using LSST data in section \ref{sec:prospects}. We combine each probe with 
the CMB distance prior calculated from the Planck 2015 result \citep{ade/etal:2015a}. The distance prior is calculated
using the shift parameter, $R_\mathrm{shift}$, and the multipole corresponding to the sound horizon at the moment of last scattering, $l_*$.
The definitions of these parameters are
\begin{equation}
\begin{aligned}
R_\mathrm{shift} &\equiv \sqrt{\Omega_m H_0^2} D_A(z_*)/c,\\
l_* &\equiv \pi \frac{D_A(z_*)}{r_s(z_*)},
\end{aligned}
\end{equation}
where $z_*=1089.94$ is the redshift of the last scattering surface, and $r_s(z_*)=144.89$ Mpc is the size of the sound horizon at redshift $z_*$.
The distance prior compresses information in the CMB power spectrum relevant for dark energy to two numbers.
\subsection{Time-delay lenses}
\label{sec:lens}
We first show how sensitive $D_{\Delta t}$ and $D_\mathrm{A}(EL)$ are
to $w_0$ and $w_\mathrm{a}$ as a function of $z_L$ and $z_S$.
\begin{figure}[t]
\begin{minipage}[b]{0.45\linewidth}
\centering
\includegraphics[width=7.8cm]{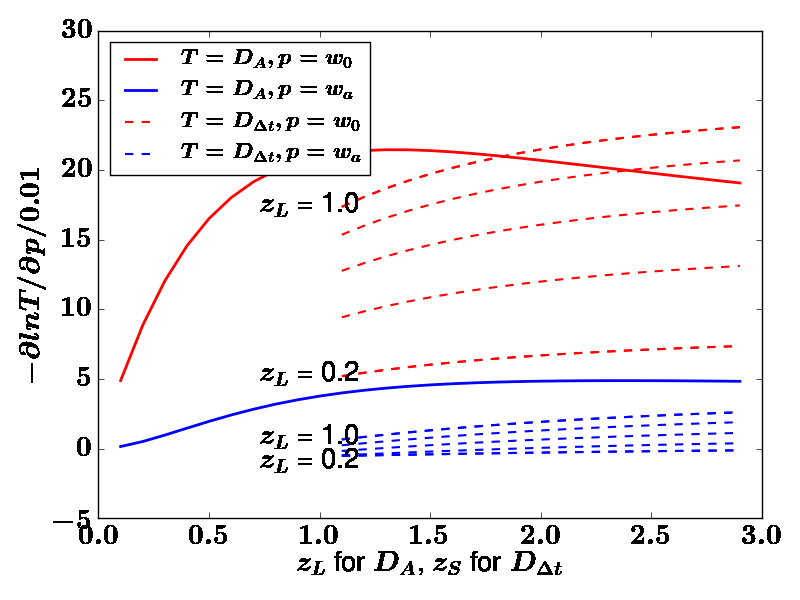}
\end{minipage}
\hspace{0.5cm}
\begin{minipage}[b]{0.45\linewidth}
\centering
\includegraphics[width=7.8cm]{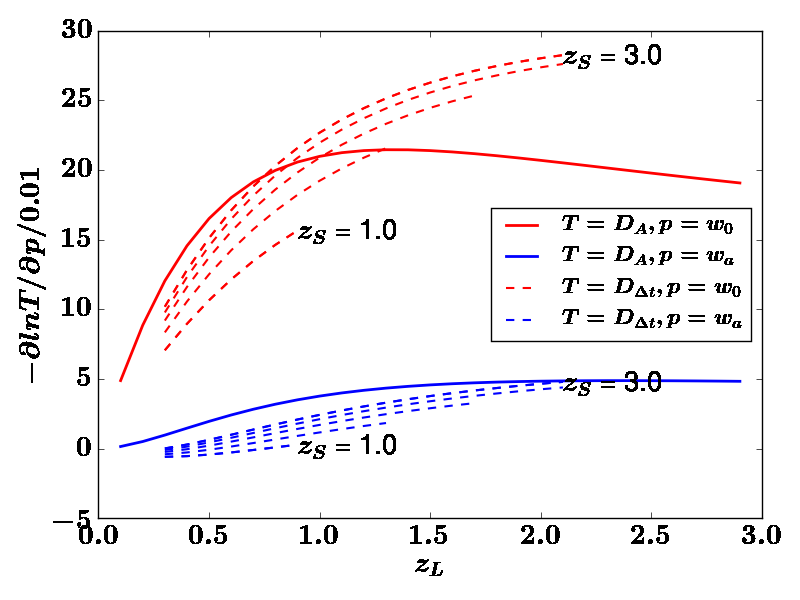}
\end{minipage}
\caption{Logarithmic derivatives of distances $T$ =($D_A$,$D_{\Delta t}$) with respect to the cosmological parameters $p$ =($w_0$,$w_a$), as a function of redshift. 
Left panel: The solid lines show $T=D_A$, while the dashed lines show $T$ =$D_{\Delta t}$,
with various combinations of the lens and the source redshift. Each dashed line corresponds to one $z_L$ in range [0.2,1.0] in increments of 0.2, and shows $-\partial \ln D_{\Delta t}/ \partial p/0.01$ as a function of $z_S$. We only show $z_S$ which is higher than the highest $z_L$ in this range. Right panel: The solid lines show  $-\partial \ln D_A/ \partial p/0.01$ as a function of $z_L$. Each dashed line corresponds to one $z_S$ in range [1.0,3.0] in increments of 0.5, and shows $-\partial \ln D_{\Delta t}/ \partial p/0.01$ as a function of $z_L$. We only show $z_L$ which is lower than the lowest $z_S$ in this range. Both panels show that $D_A$ is always more informative than $D_{\Delta t}$ for constraining $w_a$ (i.e., $\left \vert \partial \ln D_A/\partial w_a \right\vert > \left \vert \partial \ln D_{\Delta t}/\partial w_a\right \vert$), and $D_A$ is often more informative than $D_{\Delta t}$ on $w_0$.}
\label{fig:deriv}
\end{figure}

In figure \ref{fig:deriv}, we show $\partial \mathrm{ln} T/\partial p$ (where $T$ = ($D_A$,$D_{\Delta t}$) and $p$ = ($w_0$,$w_a$) ). The larger the absolute value of $\partial \ln T/ \partial p$ is, the
bigger the unmarginalized sensitivity of the distance $T$ becomes to the parameter $p$, with all the other cosmological parameters fixed at the fiducial values. Also, equation \ref{eq:fisher} shows that the information 
is proportional to $\partial T /\partial p$. The higher $z_L$ is, the more sensitive $D_{\Delta t}$ becomes to
both $w_0$ and $w_a$ for a given $z_S$, and vice versa. We find that $D_A$ is always more sensitive to $w_a$,
when all the other parameters are fixed at the fiducial values. 

\begin{figure}[H]
\begin{minipage}[b]{0.47\linewidth}
\centering
\includegraphics[height=5.8cm]{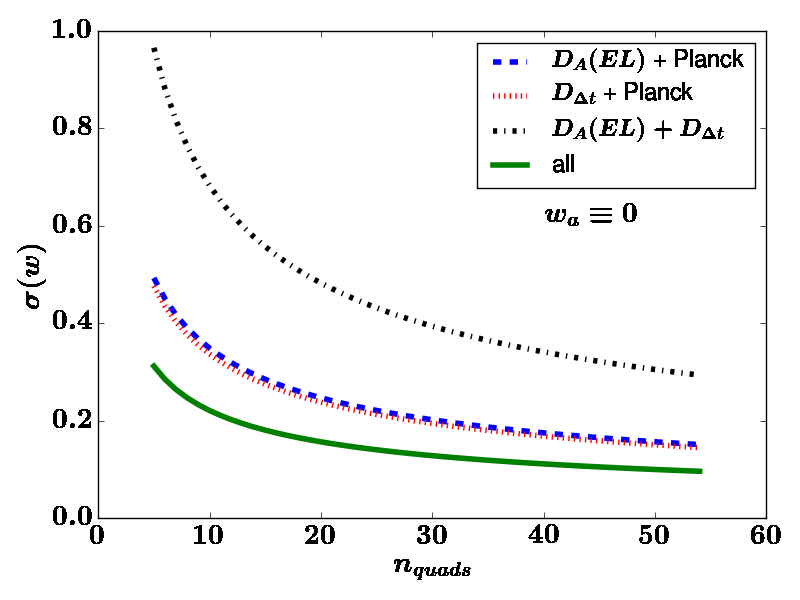}
\end{minipage}
\hspace{0.8cm}
\begin{minipage}[b]{0.47\linewidth}
\begin{center}
\includegraphics[height=5.8cm]{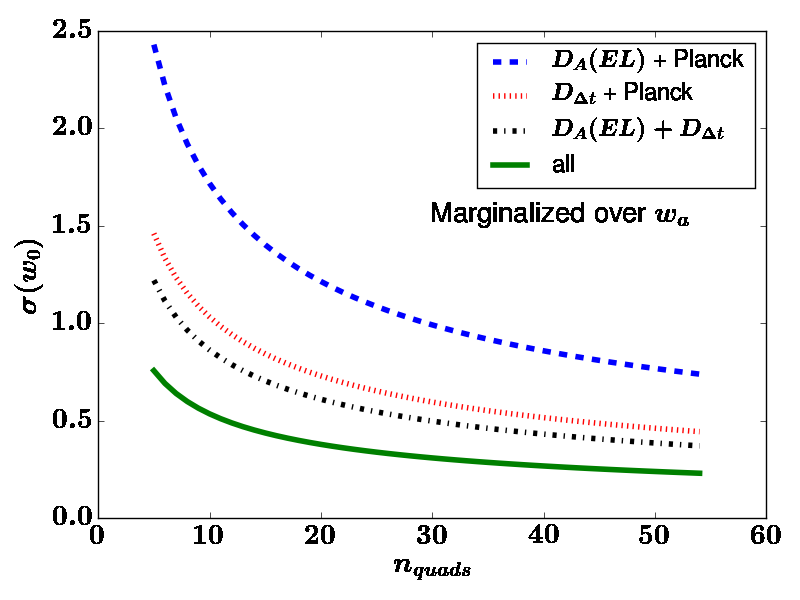}
\end{center}
\end{minipage}
\caption{The 1-$\sigma$ uncertainty in $w$, denoted as $\sigma(w)$, and that in $w_0$, denoted as $\sigma(w_0)$, from time-delay lenses as a function of the number of lenses for the left and the right panel, respectively.
The black dash-dot line is the lens-only data, while the other lines use the Planck distance priors combined with
$D_A$ (blue dashed), $D_{\Delta t}$ (red dotted), or both (green) from lenses.
(Left) o$w$CDM model.
(Right) o$w_z$CDM model marginalized over $w_a$, as well as all the other parameters.}
\label{fig:w_nquads}
\end{figure}
In figure \ref{fig:w_nquads}, we show the expected 1-$\sigma$
uncertainties in $w$ for the o$w$CDM model and $w_0$ for the o$w_z$CDM model (combined with the Planck distance priors), with all the other parameters marginalized over. 
As the Fisher matrix is proportional to $n_\mathrm{quads}$, the marginalized uncertainty in $D_A(EL)+D_{\Delta t}$ 
scales as $\propto 1/\sqrt{n_\mathrm{quads}}$. 
For o$w$CDM (left panel of figure \ref{fig:w_nquads}), the Planck distance priors combined with either $D_A$ (blue dashed line) or $D_{\Delta t}$ (red dotted)
from lenses improve the constraint on $w$ significantly compared to the lens-only case (black dash-dot). Combining all
improves the constraint further by 30\%, in comparison to that of the $D_{\Delta t}$ + Planck. 
For o$w_{z}$CDM (right panel of figure \ref{fig:w_nquads}), we find that the lens-only ($D_A$ + $D_{\Delta t}$) breaks the degeneracy between $w_0$ and the other parameters more efficiently than either combination of Planck + $D_A$ or 
Planck + $D_{\Delta t}$, yielding a tighter constraint. Combining all improves the constraint further by a factor of 2 in comparison to $D_{\Delta t}$ + Planck.

\begin{figure}[H]
\begin{minipage}[b]{0.45\linewidth}
\centering
\includegraphics[height=6cm]{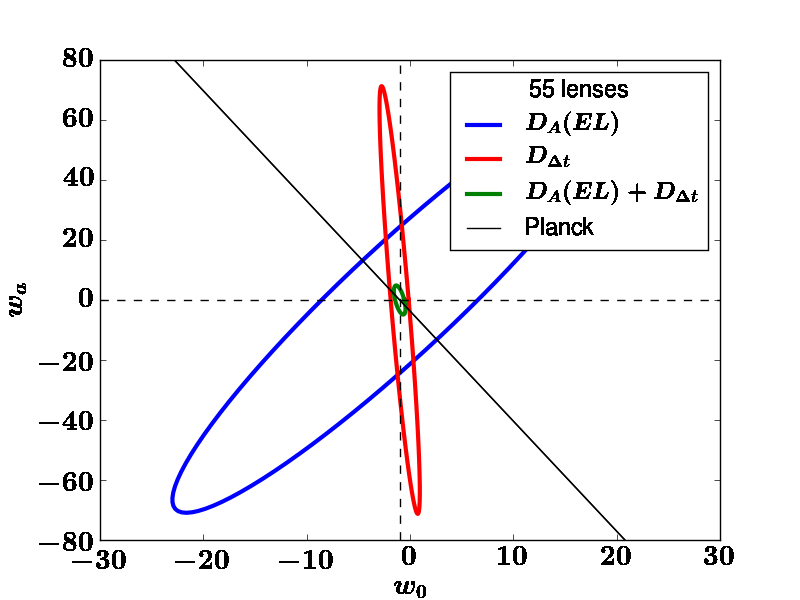}
\end{minipage}
\hspace{0.5cm}
\begin{minipage}[b]{0.45\linewidth}
\centering
\includegraphics[width=7.7cm]{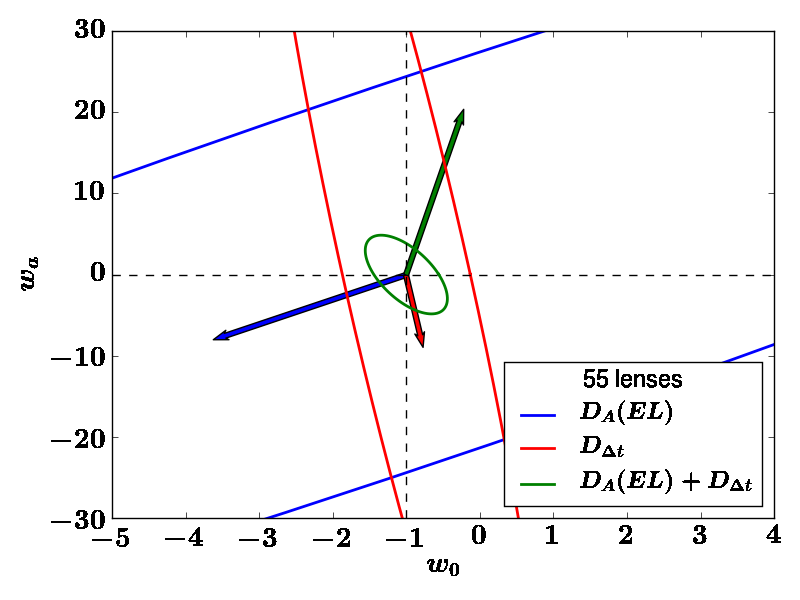}
\end{minipage}
\caption{(Left) The 68 per cent CL in the $w_0$-$w_a$ plane constrained from 
55 lenses for the o$w_{z}$CDM model.
The blue line is the marginalized constraint from
$D_A$, the red line is the marginalized constraint from $D_{\Delta t}$,
and the green line is the marginalized constraint from the combination of the $D_A$ and $D_{\Delta t}$. 
The black line is the unmarginalized constraint from Planck only. The horizontal and vertical dashed lines correspond to the fiducial values of $w_0=-1$ and $w_{a}=0$, respectively.
(Right) Zoom in of the left panel, with degeneracy directions plotted as arrows. 
The arrows show the response of $w_0$ and $w_a$ to the shift $\Delta \Omega_k = 0.02$. The blue arrow is the response for $D_A(EL)$, the red arrow is for $D_{\Delta t}$, and the green arrow is for the combination of two. The larger the arrow is relative to the contour, the more sensitive the distance is to $\Omega_k$. The alignment between the arrows and the major axis of the contours indicate that the constraints from $D_A$ (the red arrow and the red contour) and $D_{\Delta t}$ (the blue arrow and the blue contour) are individually dominated by $\Omega_k$. However, the degeneracy between $\Omega_k$ and $w$ is shown to be broken when $D_A$ and $D_{\Delta t}$ are combined. For visualization purposes, the sizes of the arrows are inflated by a factor of 100.}
\label{fig:w_wa_lens}
\end{figure}
The left panel of figure \ref{fig:w_wa_lens} is useful for understanding these results. The marginalized uncertainties in $w_a$ from either $D_A$ or $D_{\Delta t}$ individually are similar. The Planck distance prior (the black line) provides a degenerate combination of $w_0$ and $w_a$, thus cannot be marginalized. However, it is nicely orthogonal to the ones from $D_A$ (the blue contour) and $D_{\Delta t}$ (the red contour). Thus, including the combination of $D_A$ and Planck with the previous lensing constraints from $D_{\Delta t}$ reduces the uncertainty in $w_a$ significantly.
We also note that the constraint on $w_a$ from $D_A$ + $D_{\Delta t}$ (the green contour) is significantly tighter than the naive addition of the blue and the red contour, which indicates that this combination of distances effectively breaks the degeneracy between the equation of state of dark energy and the other parameters over which we marginalize.

Next, we study the degeneracy structure of parameters constrained from lensing distances, by shifting a parameter and calculating the response of the other parameters to the shift in order for the likelihood to be maximized. Specifically, we describe the degeneracy between the curvature density $\Omega_k$ and the equation of state parameters $w_0$ and $w_a$. When $\Omega_k$ is shifted by an amount $\Delta \Omega_k$, to maximize the likelihood at the new fiducial value $\Omega_k+\Delta \Omega_k$, all the other parameters have to be shifted accordingly. The general expression for the shift in an arbitrary parameter $\Delta \theta_i$ due to a shift in a fixed, single parameter $\Delta \theta_k$ that maximizes the likelihood can be calculated as 
\begin{equation}
\Delta \theta_{i} = \Delta \theta_{k}\frac{({\bf F}^{-1})_{ik}}{({\bf F}^{-1})_{kk}},
\end{equation}
where ${\bf F}^{-1}$ is the inverse of the Fisher matrix. In figure \ref{fig:w_wa_lens}, the right panel shows the projection of
the shift vectors to $w_0$-$w_a$ plane as arrows, along with the marginalized constraint contour at the fiducial parameter to display it quantitatively.
 For each of $D_A$ and $D_{\Delta t}$, the degeneracy directions are parallel to the major axes of the contours, which indicates that the degeneracies with $\Omega_k$ dominate the dependences of $w_0$ and $w_a$ to other parameters. However, when the two distances are combined, the curvature degeneracy breaks and the alignment between the error contour and the shift disappears (the green arrow and the green contour are not aligned). The relative size of the arrow to the contour shows the sensitivity of the probe to $\Omega_k$: the bigger the vector is with respect to the contour, the more sensitive the probe is to the change in $\Omega_k$. By comparing the relative size of the red and the blue arrows to the red and the blue contours, we show that $D_A$ and  $D_{\Delta t}$ are comparably sensitive to $\Delta \Omega_k$, but the combination of two increases the sensitivity significantly (the green arrow and the green contour).

\begin{figure}[H]
\begin{minipage}[b]{0.47\linewidth}
\centering
\includegraphics[height=6cm]{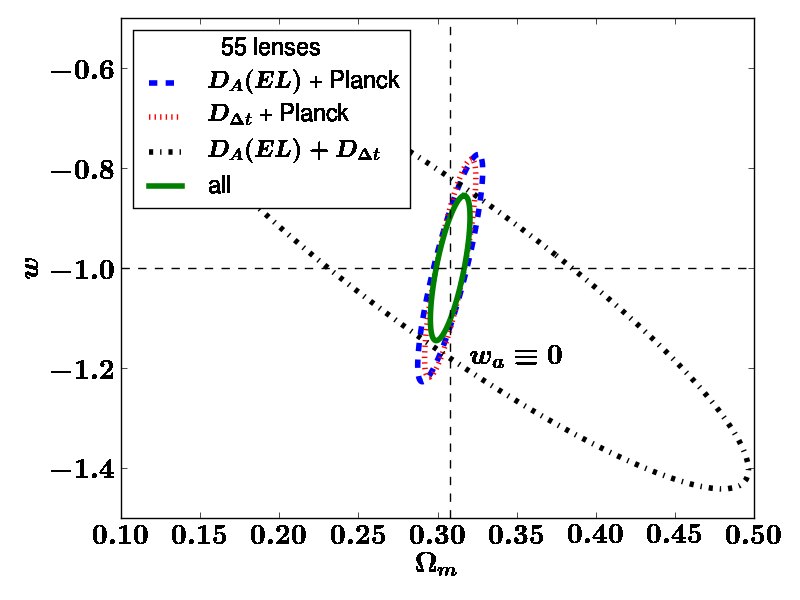}
\end{minipage}
\hspace{0.5cm}
\begin{minipage}[b]{0.47\linewidth}
\centering
\includegraphics[height=6cm]{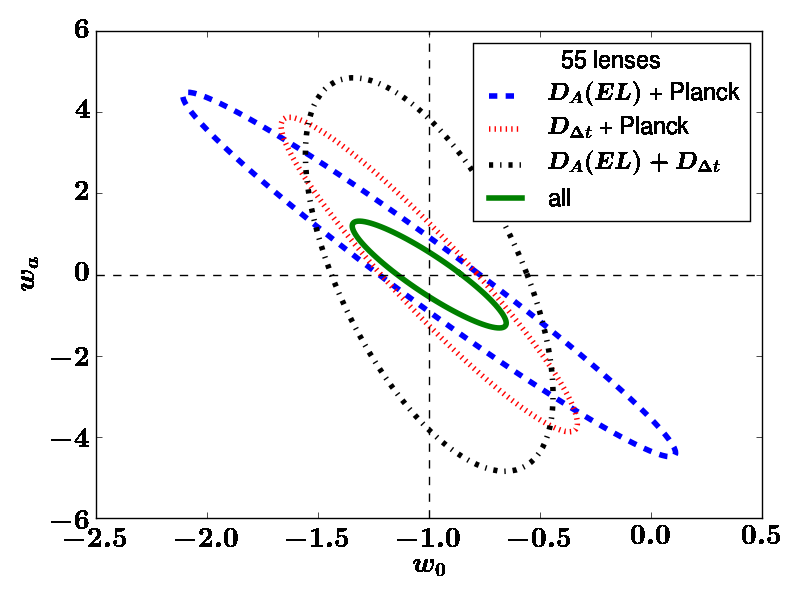}
\end{minipage}
\caption{The marginalized 68 per cent CL constraints from 55 lenses in the (left) $\Omega_{m}$-$w$ plane 
for the o$w$CDM model, and (right) $w_0$-$w_a$ plane for the o$w_{z}$CDM model.
The black dash-dot lines show the lens-only constraints from $D_A$ + $D_{\Delta t}$, the blue dashed lines the constraints from
$D_A$ + Planck, the red dotted lines the constraints from $D_{\Delta t}$ +Planck,
and the green solid lines the combination of the two distances + Planck.}
\label{fig:w_Om0_PL_lens}
\end{figure}

The left panel of figure \ref{fig:w_Om0_PL_lens} shows the joint constraints on $\Omega_m$ and $w$ for o$w$CDM. We find that the Planck distance prior plays an important role in constraining $\Omega_m$, while using both $D_A$ and $D_{\Delta t}$ combined with Planck distance prior improves the constraint on $w$ by about 30\% compared to the case of $D_{\Delta t}$ combined with Planck distance prior. The right panel of figure \ref{fig:w_Om0_PL_lens} shows the same for $w_0$ and $w_a$ for o$w_\mathrm{z}$CDM.
We also show the lensing distances + Planck constraint on $H_0$ in Appendix \ref{sec:lens_on_h}.

Next, we compare these constraints with those from Planck + BAO and Planck + SNe. We calculate the constraints from BAO and SNe using currently available data (BOSS DR11 for BAO, JLA sample for SNe). 

\subsection{BAO}
\label{sec:bao}
BOSS DR11 provides the volume-averaged distance, $D_V\equiv\left(cz(1+z)^2 D_A^2/H\right)^{1/3}$, 
at two effective redshifts (0.32, 0.57) obtained from the BAO peak position in the spherically averaged two-point 
functions. 
The lower redshift is the LOWZ sample, and the higher redshift is the CMASS sample.
Also, by separately measuring the two point functions along the line of sight and the direction perpendicular to it, 
the DR11 CMASS sample separately constrains the angular diameter distance $D_A$ and the 
Hubble parameter $H$ at $z= 0.57$ \citep{anderson/etal:2013, sanchez/etal:2013}. 
To account for the correlation among $D_V$, $D_A$ and $H$, we use the full likelihood of the 
CMASS sample for the analysis provided by the BOSS collaboration \citep{anderson/etal:2013}.

In our analysis we assume that the sound horizon scale at the baryon drag epoch, $r_{s,drag}$, 
is fixed as $r_{s,drag} = 149.28 ~\mathrm{Mpc}$ \citep{anderson/etal:2013}. 
We then combine the cosmological constraints from $D_V$ at $z=0.32$, and 
$D_A$ and $H$ at $z = 0.57$ with the Planck distance prior. 
We calculate the Fisher matrix by taking the derivatives of the log likelihood at the fiducial cosmology.
The results are shown in figure \ref{fig:w_Om0}.
The precision of the BAO data yields the narrowest contours on the $\Omega_m$-$w$ (for o$w$CDM) and $w_0$-$w_a$ (for
o$w_\mathrm{z}$CDM) planes. However, due to the limited number of redshifts ($z=0.35$ and $z=0.57$),
the degeneracy is not broken efficiently; thus, the expected Planck + lens from 55 lenses can
improve the constraints significantly, even though the precision of lensing data per redshift is not as precise as BAO.

\subsection{SNe}
\label{sec:sne}
We now study the constraints from Planck + SNe.
We use the JLA data \citep{betoule/etal:2014} to calculate the constraints from SNe. 
JLA uses Supernovae Legacy Survey (SNLS), Sloan Digital Sky Suvey-II (SDSS-II) Supernova survey and a few low-redshift samples. The redshift of subsamples are: 
the low-redshift sample ($z<0.1$), SDSS-II ($0.05<z<0.4$), and 
SNLS ($0.2<z<1$). There are 740 spectroscopically confirmed type Ia SNe in JLA.
SDSS-II is used for anchoring the distances, 
and also an empirical relation between the host galaxies and the supernovae brightness is used as an extra calibration 
for the absolute magnitude of the SNe. 
For the calibration, there are 4 additional nuisance parameters that are taken into account in JLA:
$\alpha$, which scales the stretch of the light curve in time-domain; $\beta$, which scales the color
at the peak of the light curve; $M$, which is the absolute B band magnitude of the SNe at the peak of the light curve; and $\Delta_M$, which characterizes the peak absolute magnitude change due the stellar mass of the
host galaxy. 

\begin{figure}[H]
\begin{minipage}[b]{0.47\linewidth}
\centering
\includegraphics[height=5.6cm]{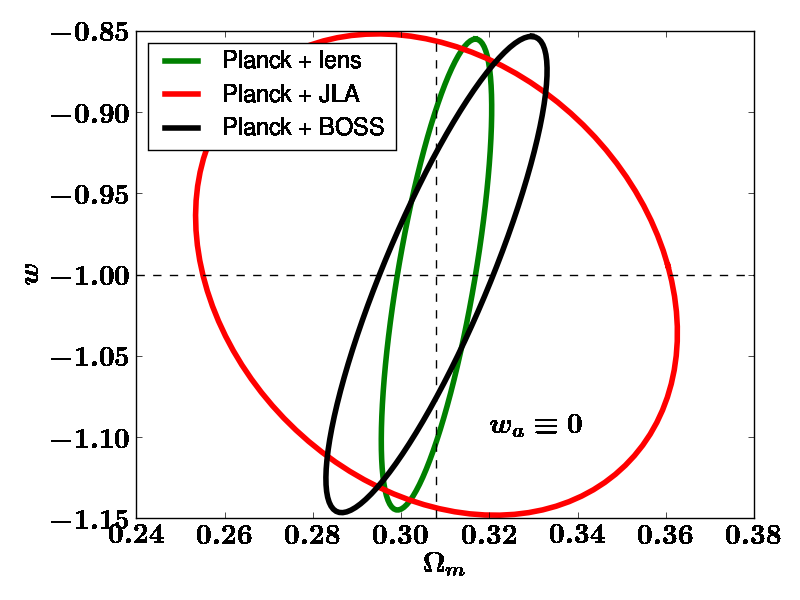}
\end{minipage}
\hspace{0.5cm}
\begin{minipage}[b]{0.47\linewidth}
\centering
\includegraphics[height=5.6cm]{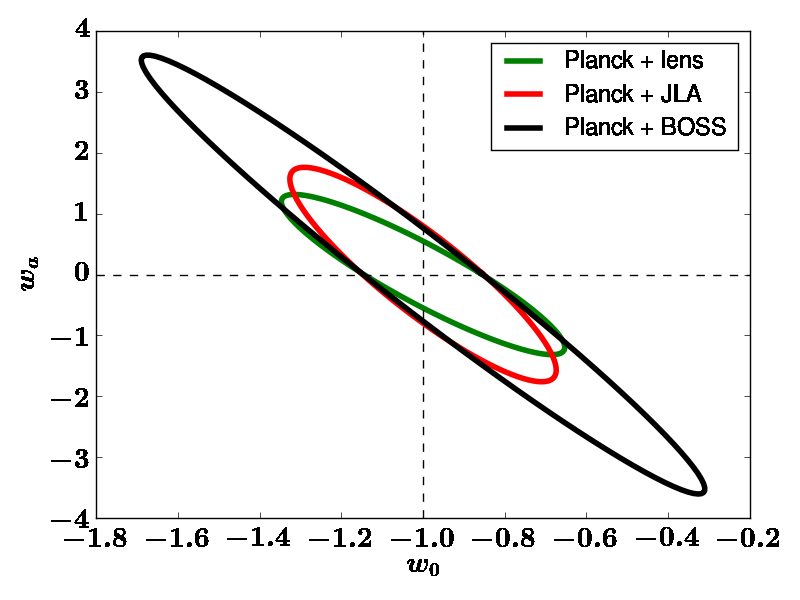}
\end{minipage}
\caption{The marginalized 68 per cent CL constraints from strong lenses, SNe, and BAO, each combined with the Planck distance prior, in the
(left) $\Omega_{m}$-$w$ plane for the o$w$CDM model, and (right) $w_0$-$w_a$ for the o$w_{z}$CDM model.
The green lines show the constraints from Planck + lens, the red lines Planck + JLA, and the black lines 
Planck + BOSS. }
\label{fig:w_Om0}
\end{figure}

We use Montepython \citep{audren/etal:2012} to sample the JLA likelihood.
Specifically, we run Markov Chain Monte Carlo to sample the likelihood surface, and compute the covariance matrix in the cosmological parameters. We then use its inverse as the JLA Fisher matrix.
The results are shown in figure \ref{fig:w_Om0}.
While the absolute distances, such as those from BAO and lenses, are effective at measuring $\Omega_k$
when combined with CMB \citep{knox:2005}, the relative distances from SNe are not. Thus, when $\Omega_k$
is set free, the constraints on $\Omega_m$ for o$w$CDM
from 55 lenses combined with Planck are significantly better than those from 740 SNe combined with Planck.

\subsection{Comparison to future BAO and SNe predictions}
\label{sec:prospects}
 With several billion galaxies expected to be detected with LSST, BAO will allow measurements of distances with $\sim$2\% precision in the redshift range $1<z<3$ \cite{zhan/etal:2008}. Combined with Planck, BAO will constrain $w_0$ with uncertainty $\sim$0.4 and $w_a$ with $\sim$1 \cite{abate/etal:2012}. Also, 500,000 SNe are expected to be detected in 10 years of LSST operation in the redshift range $0.1<z<1.2$. With a subsample of 50,000 SNe only, the data will constrain $w_0$ with uncertainty $\sim$0.05,
and $w_a$ to order unity, assuming a flat universe \cite{abell/etal:2009}; in combination with Planck, the full sample of SNe constraints will be $\sim$0.25 for $w_0$ and $\sim$1.2 for $w_a$ for the o$w_z$CDM model \cite{abate/etal:2012}. We note that a modest sample of 55 lenses combined with the Planck distance prior constrains $w_0$ and $w_a$ to $\sim$0.4 and $\sim$1.2, respectively (see, e.g., Figure 6), which is comparable in precision to those expected from future BAO or SNe samples in the LSST era, when each is combined with Planck. Therefore, strong lenses provide an independent and competitive probe of dark energy.  Needless to say, lensing, SNe, and BAO are affected by different systematic errors, and thus cross-checking the results using these three low-redshift probes of the expansion of the universe will be powerful.
\section{Pivot redshift}
\label{sec:pivot}
The equation of state of dark energy, $w(z)$, can be re-written as 
\begin{equation}
w=w_0+(1-a)w_a = w_p+(a_p-a)w_a,
\end{equation}
where $w_p\equiv w_0+(1-a_p)w_a$ \citep{huterer/turner:2001}. 
In this parameterization, the \textit{pivot redshift} $z_p = \frac{1}{a_p}-1$ is defined as the redshift where 
the uncertainty in $w$ is minimized. The uncertainty in $w_p$ shows how well a probe can measure 
the equation of state $w$, as $w_p$ is orthogonal to $w_a$ by construction, and thus is not coupled to the time variation of
$w$ \citep{huterer/turner:2001,albrecht/etal:2009}. 
$z_p$ shows at which redshift the main constraint on $w$ is coming from: this pivot redshift varies depending on the probes, their redshift distributions and the measurement uncertainties, and can be negative.
\begin{figure}[H]
\begin{center}
\includegraphics[height=8cm]{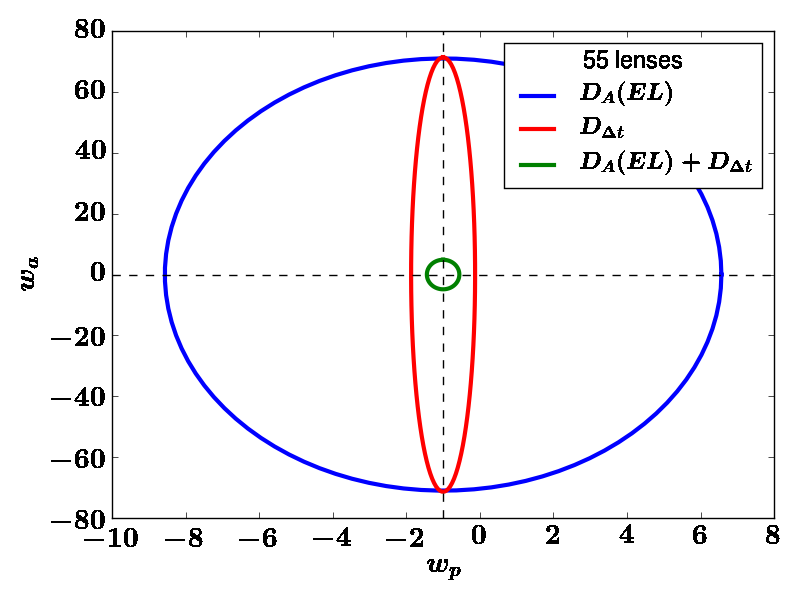}
\caption{Same as figure \ref{fig:w_wa_lens}, but in $w_p$-$w_a$ plane. $z_p$ is calculated separately for each probe: $z_p(D_A) = -0.226$, $z_p(D_{\Delta t}) = 0.0251$, and $z_p(D_A +D_{\Delta t}) = 0.0759$.}
\label{fig:wp_wa_lens}
\end{center}
\end{figure}
Figure \ref{fig:wp_wa_lens} shows the constraints on ($w_p$,$w_a$) using the lens distances alone in 
$w_p$-$w_a$ plane. The constraint on $w_p$ from $D_A$ alone is much weaker than that from $D_{\Delta t}$, while the constraints on $w_a$ from both distances are comparable. However, by combining the two distance measures, the constraint on $w_p$ improves by a factor of 2, and that on $w_a$ improves by a factor of 24, due to complementary degeneracy directions as shown in figure \ref{fig:w_wa_lens}.

The pivot redshift, $z_p$, and the uncertainties in $w_p$ and $w_a$ for two combinations of probes (Planck + BOSS + JLA, and 55 lenses (denoted by “Lens”) + Planck + BOSS + JLA) for three different cosmological cases ((1) $\Omega_k=0$ and $\Omega_m = 0.309$, (2) $\Omega_k=0$ and marginalized over $\Omega_m$, and (3) marginalized over both $\Omega_k$ and $\Omega_m$) are summarized in table \ref{tb:pivot}.
\begin{table}
\centering
\begin{tabular}{cccccc}
  \toprule[1.5pt]
  \head{Cosmological model}&\head{Probe} & \head{$z_p$} & \head{$\sigma(w_p)$} & \head{$\sigma(w_a)$} & \head{FoM}\\
    \midrule
  \multirow{2}{*}{Flat, fixed $\Omega_m$} &CMB+BAO+SNe & 0.288 & 0.0375 & 0.364 &73.3\\
  &Lens+CMB+BAO+SNe & 0.487 & 0.0296 & 0.166 &204\\
  \midrule
  \multirow{2}{*}{Flat, marginalized over $\Omega_m$} &CMB+BAO+SNe & 0.358 & 0.0487 & 0.413 & 49.7\\
  &Lens+CMB+BAO+SNe & 0.386 & 0.0310 & 0.363&88.9\\
   \midrule
  \multirow{2}{*}{\makecell{Marginalized over $\Omega_m$ and $\Omega_k$\\(o$w_z$CDM)}}
  &CMB+BAO+SNe & 0.215 & 0.0625 & 1.03 &15.5\\
  &Lens+CMB+BAO+SNe & 0.245 & 0.0479 & 0.621&33.6\\
  \bottomrule[1.5pt]
\end{tabular}
\caption{Pivot redshift, $z_p$, uncertainties in $w_p$ and $w_a$, and the Figure of Merit (FoM) for Planck + BOSS + JLA and Lens + Planck + BOSS + JLA. We test three cases: (1) $\Omega_k=0$ and $\Omega_m = 0.309$, (2) $\Omega_k=0$ and marginalized over $\Omega_m$, and (3) marginalized over both $\Omega_k$ and $\Omega_m$.
As there is no correlation between $w_p$ and $w_a$, FoM $= 1/(\sigma(w_p) \sigma(w_a))$.}
\label{tb:pivot}
\end{table}
In comparison to Planck + BOSS + JLA, adding 55 lenses tightens the FoM by a factor of 2.78, 1.62 and 2.32 for the three models, respectively. In particular, when $\Omega_k$ is allowed to vary, the uncertainty in $w_a$ is reduced by almost a factor of 2 by including the lenses, which shows that the combination of $D_A$ and $D_{\Delta t}$ is powerful in breaking the $\Omega_k$-$w$ degeneracy. This is consistent with our argument
in section \ref{sec:lens}. Also $z_p$ becomes higher as we include the lens distances in every case, which is typically beneficial when combining these probes of geometry with probes of the growth of cosmic structure.

\section{Conclusion}
\label{sec:conclusion}
There is more valuable cosmological information in the strongly lensed systems
than measurements of the Hubble constant from time delays. In this paper, we have demonstrated that the
addition of the angular diameter distance measurements to the quantity ($D_{\Delta t}$) that
captures the cosmological information from time delays in the same sample of
lenses provides crucial help in breaking cosmological parameter
degeneracies. This improvement is most significant in some of the most
interesting parameterizations that are currently being studied, such as when
curvature of the universe and the time-variation in the equation of state of
dark energy are allowed to be nonzero.

We have calculated the lensing constraints based on the predictions for the
LSST survey, adopting a catalog of 55 quadruply imaged lenses (out of a much
larger total number) that should have sufficiently good information that all
observable quantities of interest in the lenses can be accurately measured. We
have combined the forecasted lensing information from both the angular
diameter distance and the time-delay distance. We then compared this lensing
constraints with that from the BOSS DR11 and from the JLA type Ia supernova sample --
each combined with the Planck 2015 distance prior.

\begin{figure}[H]
\begin{minipage}[b]{0.47\linewidth}
\centering
\includegraphics[height=6cm]{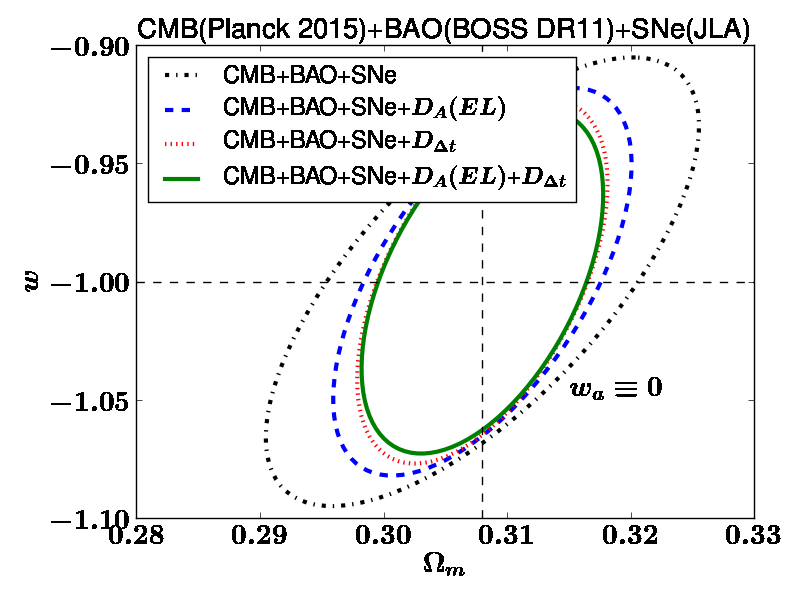}
\end{minipage}
\hspace{0.5cm}
\begin{minipage}[b]{0.47\linewidth}
\centering
\includegraphics[height=6cm]{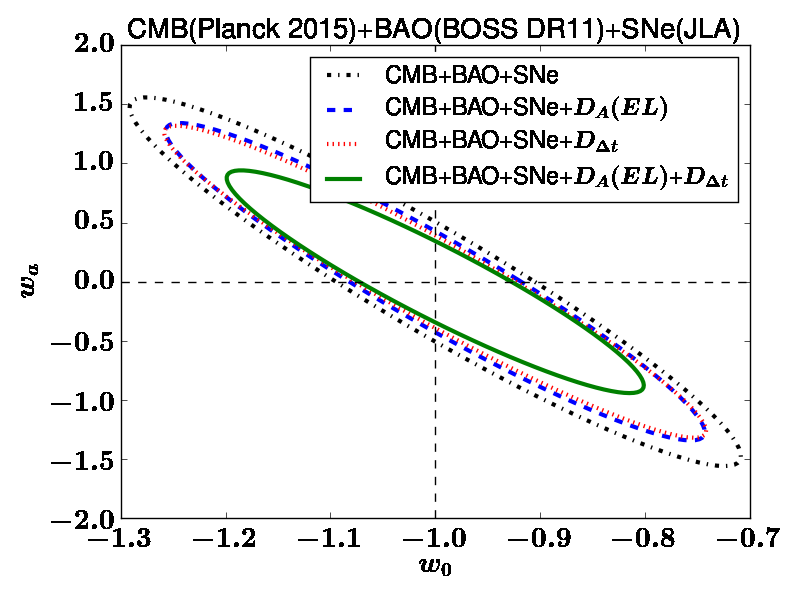}
\end{minipage}
\caption{The marginalized 68 per cent CL from strong lenses, combined with Planck, BAO and SNe in the
(left) $\Omega_{m}$-$w$ plane for the o$w$CDM model, (right) $w_0$-$w_a$ for the o$w_\mathrm{z}$CDM model.}
\label{fig:w_Om0_BOSS_LSST_PL_JLA}
\end{figure}

We find that the combined lensing information significantly helps constrain
the cosmological parameters, particularly when curvature is allowed to vary
and when the equation of state of dark energy is allowed to be
time-dependent. For example, lensing information would improve the current
BAO+CMB+SN constraints on $w_a$ by a factor of two, and those on the overall
figure of merit of dark energy by about a factor of two relative to the
case with no lensing (see Table \ref{tb:pivot} and figure \ref{fig:w_Om0_BOSS_LSST_PL_JLA}). 
Key to this significant
improvement is lensing's ability to break the degeneracy between curvature and the equation
of state parameters; see figure \ref{fig:w_wa_lens}.

We are therefore very optimistic about the prospects of a select, accurately
observed subsample of strong gravitational lenses to improve our constraints
on dark energy. Fortunately, the lensing samples are a guaranteed product of
the current and upcoming wide-field, deep surveys such as HSC, DES and LSST.

\acknowledgments
We thank Stefan Hilbert, Phil Marshall, Jan Grieb, and Ariel S\'{a}nchez for providing helpful information on the analysis. 
We thank Eric Linder for helpful comments, which helped identify an error in our Fisher matrix calculation for $\Omega_{m}$ and $\Omega_k$ in the first draft of the paper. We also thank Peter Schneider and Phil Marshall for the comments and the discussions for the paper. DH has been supported by NSF under contract AST-0807564 and DOE under
contract DE-FG02-95ER40899, and also by the DFG cluster of excellence ``Origin
and Structure of the Universe'' (\url{www.universe-cluster.de}).

\bibliography{cosmography_predictions_160126}
\appendix
\section{Lensing constraints on $H_0$}
\label{sec:lens_on_h}
\subsection{$H_0$ in o$w$CDM and o$w_z$CDM models}
We show the constraints on $H_0$ from $D_A$ and $D_{\Delta t}$ combined with the Planck distance prior. 
\begin{figure}[H]
\begin{minipage}[b]{0.47\linewidth}
\centering
\includegraphics[height=6cm]{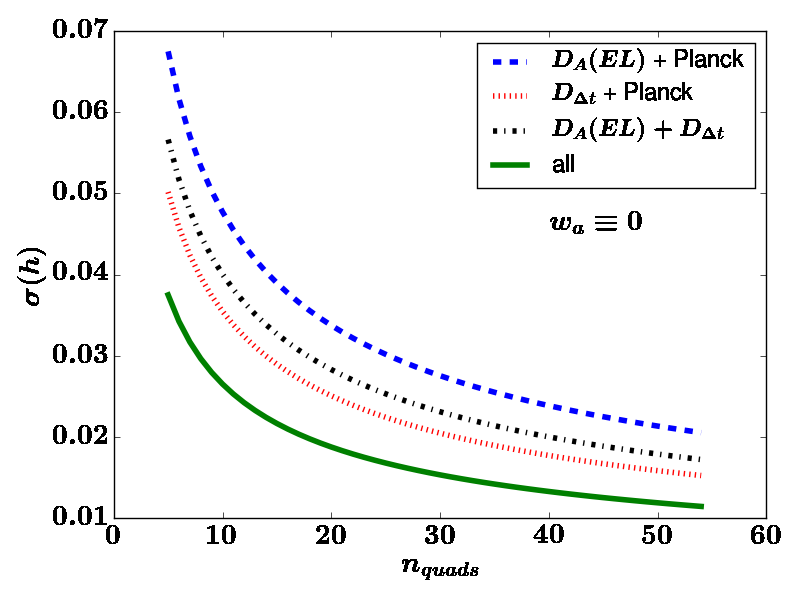}
\end{minipage}
\hspace{0.8cm}
\begin{minipage}[b]{0.47\linewidth}
\begin{center}
\includegraphics[height=6cm]{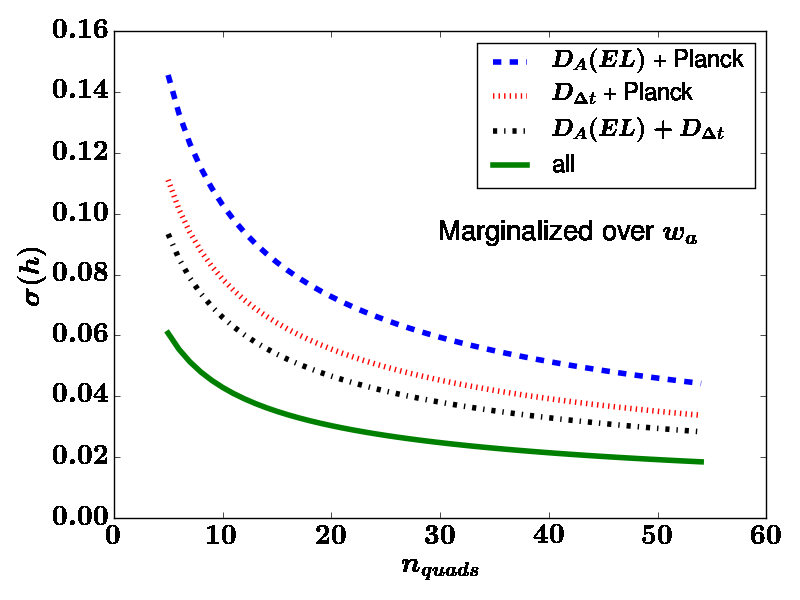}
\end{center}
\end{minipage}
\caption{The 1-$\sigma$ uncertainty in $h = H_0/100$ km/s/Mpc from time-delay lenses as a function of the number of quadruply imaged lenses for the (left) o$w$CDM model, and (right) o$w_z$CDM model.}
\label{fig:h_nquads}
\end{figure}
Figure \ref{fig:h_nquads} shows the expected 1-$\sigma$ uncertainties in $h$ from strong lenses 
combined with the Planck distance priors. As $D_{\Delta t}$ is mostly sensitive to $H_0$, the constraining power
of $D_{\Delta t}$ + Planck (red dotted line) is more powerful than that of $D_A$ + Planck (blue dashed line). When $w$ is fixed as a constant (o$w$CDM model, left panel), $D_{\Delta t}$ + Planck are more powerful than $D_{\Delta t}$ + $D_A$ (black dot-dashed line). When $w$ is allowed to vary (o$w_z$CDM model, right panel), however, $D_{\Delta t}$ + $D_A$ is more powerful than $D_{\Delta t}$ + Planck. This is due to the degeneracies between $H_0$, $\Omega_k$ and $w$ from the linear CMB  constraints alone \citep{efstathiou:1998}, which cannot be broken by $D_{\Delta t}$. However, ref. \citep{hu:2005} has shown that the main degeneracy from CMB constraints is between $w$ and $H_0$, and as shown in section \ref{sec:lens}, the combination of lensing distances is powerful in breaking the degeneracy between $\Omega_k$ and $w$.
Thus, the combination of Planck and the lensing distances shows 30\% improvement in constraining $h$. 
\subsection{$H_0$ in flat $\Lambda$CDM model}
\label{sec:h_flat_univ}
\begin{figure}[H]
\begin{center}
\includegraphics[height=6cm]{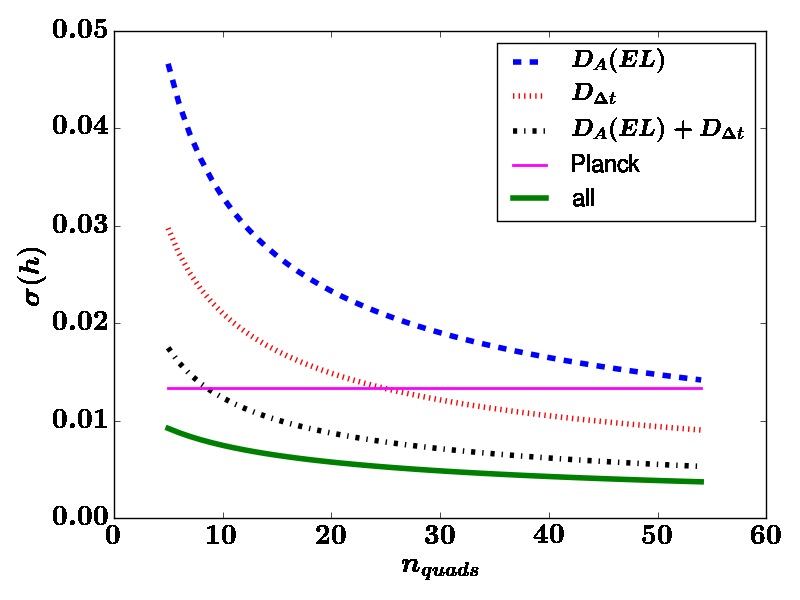}
\caption{Same as figure \ref{fig:h_nquads}, but for the flat $\Lambda$CDM model. Here the blue dashed, red dotted and black dash-dot lines are from the lensing distances alone, not combined with Planck. We show the constraints from Planck as the horizontal magenta solid line and Planck + lensing distances as the green solid line. Planck-precision constraint in $h$ is achievable with 10 lenses when we use both $D_A$ and $D_{\Delta t}$, while we need 25 lenses to achieve the same constraint from $D_{\Delta t}$ alone. }
\label{fig:flat_h_nquads}
\end{center}
\end{figure}

We show the constraints on $H_0$ for the $\Lambda$CDM model in figure \ref{fig:flat_h_nquads}. 
Assuming that 5\% precision measurements in individual distance (both $D_A$ and $D_{\Delta t}$) are achievable from lens systems, 10 lenses are enough to measure the Hubble constant to the same precision as Planck. The number of required lenses to achieve the same precision increases to 25 if constraints are from $D_{\Delta t}$ only. 
\section{Constraints assuming the flat universe}
\label{sec:flat_univ}
\begin{figure}[H]
\begin{minipage}[t]{0.45\linewidth}
\centering
\includegraphics[height=6cm]{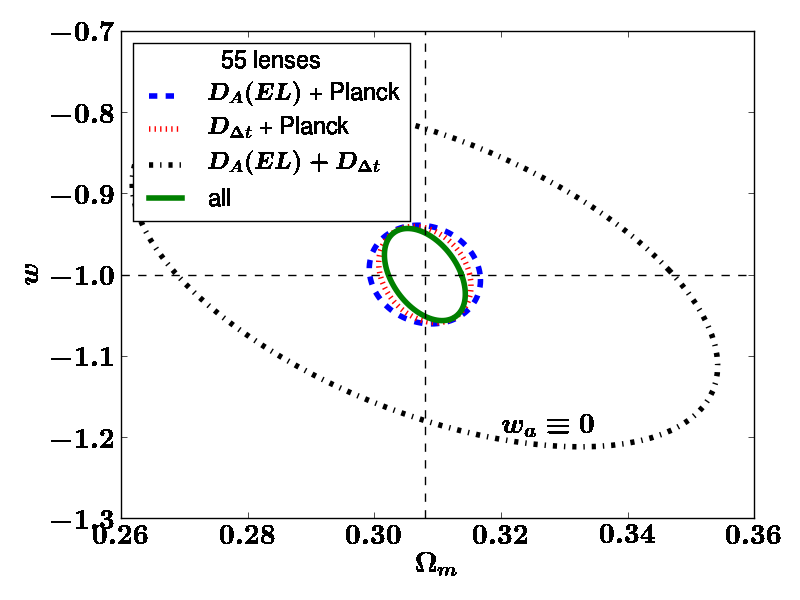}
\end{minipage}
\hspace{0.8cm}
\begin{minipage}[t]{0.45\linewidth}
\centering
\includegraphics[height=6cm]{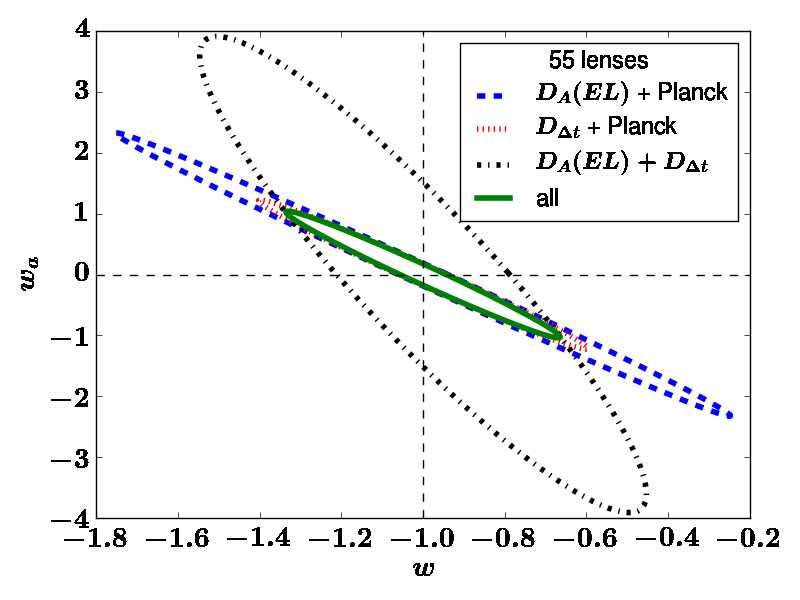}
\end{minipage}
\caption{Same as figure \ref{fig:w_Om0_PL_lens}, but for the flat (left) $w$CDM and (right) $w_z$CDM model.}
\label{fig:flat_w_Om0_PL_lens}
\end{figure}

\begin{figure}[H]
\begin{minipage}[t]{0.45\linewidth}
\centering
\includegraphics[height=6cm]{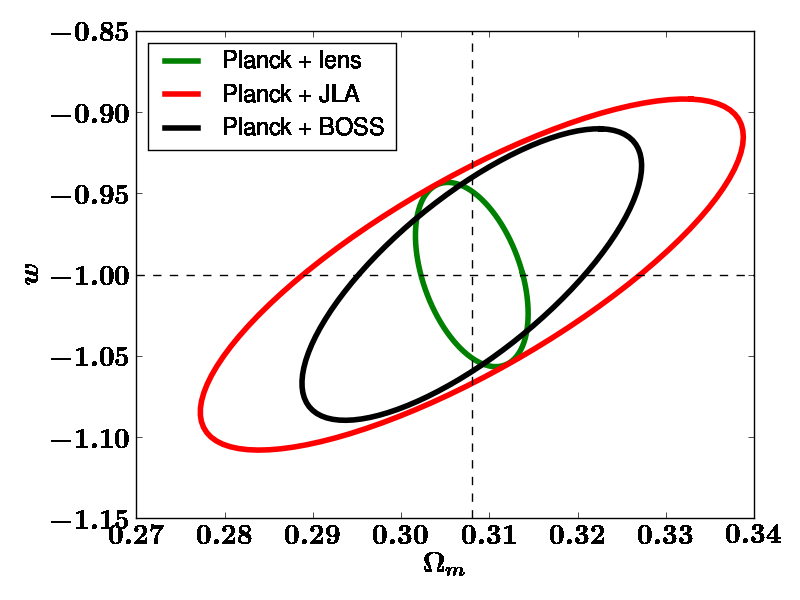}
\end{minipage}
\hspace{0.8cm}
\begin{minipage}[t]{0.45\linewidth}
\centering
\includegraphics[height=6cm]{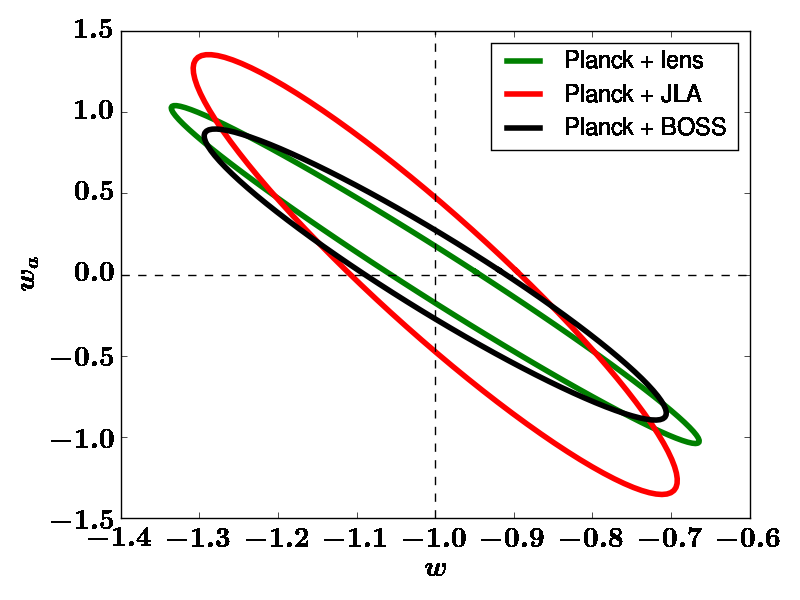}
\end{minipage}
\caption{Same as figure \ref{fig:w_Om0}, but for the flat (left) $w$CDM and (right) $w_z$CDM model.}
\label{fig:flat_w_Om0}
\end{figure}

In section \ref{sec:lens}, we have shown that the lensing distances are powerful probes for the curvature of the universe.
Specifically, $D_A$ and $D_{\Delta t}$ respond to curvature differently in $w_0$-$w_a$ plane, thus the combination of two
gives a strong constraint on $\Omega_k$.
We repeat the same analysis for the flat universe model ($\Omega_k \equiv 0$).
The model parameters are summarized as
%
\begin{equation}
\vec{\theta} \in \{\Omega_m, w, h\} \quad \mbox{(flat $w$CDM model)},
\end{equation}
and
\begin{equation}
\vec{\theta} \in \{\Omega_m, w_0, w_a, h\} \quad \mbox{(flat $w_z$CDM model)}.
\end{equation}
The constraining contours for these models are shown in figures \ref{fig:flat_w_Om0_PL_lens} and \ref{fig:flat_w_Om0}.


Figure \ref{fig:flat_w_Om0_PL_lens} shows that under the flatness assumption, the constraints from $D_{\Delta t}$ + Planck are already as tight as those from $D_A + D_{\Delta t}$ + Planck, i.e., the constraining power from $D_A$ in flat universe is minor.
Figure \ref{fig:flat_w_Om0} shows that the 55 lenses combined with Planck still constrain the equation of state better
as compared to Planck + JLA and Planck + BOSS for the flat $w$CDM model (left panel), and comparably well as Planck + BOSS for the flat $w_z$CDM model (right panel).



%
%
%
%
%
%
%
%

\end{document}